\documentclass[11pt]{article}
\usepackage{float}
\usepackage{array}
\usepackage{cool}
\usepackage{epstopdf}
\usepackage{epsfig}
\usepackage{bbold}
\usepackage{eulervm}
\usepackage{hyperref}
\usepackage{tabularx}
\hypersetup{colorlinks=true,linkcolor=blue,citecolor=blue}
\let\oldmb\mathbold
\protected\def\mathbold{\oldmb}
\usepackage{amsmath}
\usepackage{graphicx}
\usepackage{color}
\begin{document}
\title{\bf Photonic Hall effect }

\author{D. Jahani\thanks { e-mail address:  dariush.jahnai@gmail.com}\ ,  A. Alidoust Ghatar , L. Abaspour, T. Jahani }
 \maketitle {\it \centerline{  \emph{}}\maketitle {\it \centerline{\emph{Institute of solid states physics, Leibniz University of Hannover, Appelstr. 2, D-30167 Hannover, Germany }}

\begin{abstract}
\emph{{In this work, we report on the emergence of a photonic Hall effect (PHE) system within a narrow filtered background of a one-dimensional defective optical dielectric structure with graphene under the static QHE regime. It is observed that at low temperature and relatively strong applied magnetic fields, electromagnetic defective transmission spectra corresponding to the two decoupled right- and left-handed polarized modes possess a step-like transmission feature which are referred to as "quantum Hall defect modes" (QHD modes or QHDs) in this paper. Tunable growing transitional transmission steps for QHDs with increasing the magnetic field intensity was shown to be possible. Observation of sensitive magneto-transmission oscillations to the thermal excitations in the last plateaus with slow ascending toward unity is another special feature noted in this work. The results of this study which is carried out based on a rapid standard calculations for transfer matrix approach is supplied with commercial simulations marking the first PHE  system promise an proper candidate for new photonic applications, especially new tunable magneto-based lenses and photonic magneto-thermal sensors.}}
\end{abstract}
\vspace{0.5cm} {\it \emph{Keywords}}: \emph{Photonic Hall effect; Quantum Hall defect modes; Graphene; Photonic structures.
 }
\section{Introduction}
 \emph{Recently some attentions have been paid to optical quantum Hall effect (QHE) systems in ac regime in graphene, a newly well-known 2D monolayer of carbon atoms with low energy excitations governed by massless Dirac equation [1-4]. While the focus of these works are on the optical properties of QHE systems, so far the emergence of photonic Hall effect (PHE) for the transport of electromagnetic (EM) waves analogues to the electronic Hall plateaus seen as a function of the applied magnetic field has remained uncovered. The absence of this PHE system is felled in photonics since the static QHE which itself emerges in presence of a pair of crossed static electric and magnetic fields applied on a 2D electronic gas system has provided an excellent route toward understanding the underlying electronic characteristic of charge carriers in the matter. In fact, it was the static QHE measurement in graphene that for the first time proved its massless Dirac electronic nature in the year 2005 after which huge increasing attentions were drawn to its peculiar properties [5]. We note that apart from the expected quantization version in QHE, it is the existence of plateaus due to Anderson localizations in the presence of the disorder that has made electronic QHE one of the most interesting effects in the electronic systems [6,7]. Now, EM waves transport through periodic optical structures with the stacks of periodic dielectric material for which introducing disorder as a defect layer is also possible has shown to possess similar properties analogues to the transport prosperities that electronic waves in a semiconductor exhibit [8,9]. Although Anderson localization as well as some other features for photons i.e, the non-interacting bosoninc gas, has also been pointed in optical communications [10], the observation of photonic Hall effect (PHE) still could be an issue to be most sought after in optics.}
\par
\emph{In optics, the propagation of EM waves through periodic optical structures which are almost building blocks for all optical applications demonstrates a certain forbidden bandgap (PBG) suggesting a great potential for optoelectronics applications in optical devices [11-,12]. However, the existence of a single frequency mode within these forbidden zones by introducing disorder, in order to split the periodicity of the bands analogous to creating a defect mode in the forbidden energy region of doped semiconductors-which could be even more applicable than their pure counterparts-offers this desirable potential to engineer the EM radiation in a more controllable manner [13,14]. Controlling the properties of these single modes in order to extend the real world applications of photonic structures in the presence of some well-known parameters such as electric and magnetic fields has recently attracted considerable attentions [15,18]. Now, the peculiar properties of graphene under the effect of a magnetic field which develop an unconventional quantum Hall effect allow us to consider this one-atom thick carbonic material in a disordered one-dimensional optical structure to form a defective photonic Hall system by introducing graphene in Hall situation as a covering material for the defect layer of the defective structure. We note that applying graphene also as an embedded material with low loss propagation of EM waves especially in infrared region of the spectrum in photonic structures was first suggested by Berman et. al [19,20]. Incorporating graphene's capability in having tunable PBGs in 1D Pshotonic structures into the need to tune the EM defect mode is a recent growing topic of research and is of great importance for our study with the aim of founding a truly PHE system with the peculiar tunable properties in optics.}
 \par
 \emph{ The main aim of the present work is to study the behavior of the transmission of QHDs to show that our suggested structure develops a PHE system with peculiar properties in optics and optoelectronics. Along our main aim is to show that the left-and right-handed polarized QHDs in the normal propagation are highly tunable and show different transmission modes in  the frequency space in terms of the intensity of the magnetic field. Similar to the behavior of the most of the electronic QHE systems, our sample shows peculiar behavior for different temperatures and chemical potential which are discussed completely. Finally, it is also worth mentioning that the proposed structure in this paper can open a new window toward investigations of the defective photonc structures which are taken rather less attention in the high level optics technology.    }\par
 \emph{The reader could find the remaining discussions of this relatively rapid communications into the following 2 sections: In section 2, the underlying theory behind the model is discussed. Along with the underlying theory, the numerical simulations of the transmission properties of our 1D PHE structure for the most general situations are followed. Finally, in section 3, the general remarks are provided.}
\section{Formalism and simulations}\label{section.two}
\emph{ In this section, we first try to develop a relatively rapid but manageable procedure based on which rather tedious formalism for the propagation matrix of EM wave transport in a single layer graphene (SLG) under Hall regime in the presence of dielectrics would be obtained. We then expand our approach to the mentioned 1D photonic model in order to numerically simulate the transmission properties of the proposed structure. In this regard, we consider a defect layer to be composed of epitaxially grown graphene on silicon carbide ($SiC$) with the thickness, $d$, and dielectric constant of $4.4$ which has shown to be more suitable for magneto-optical studies because of the well-controlled morphology and essentially unlimited size [1-3]. Therefore, with $A$ ($Si$) and $B$ ($SiO_{2}$) representing the dielectric materials, and D the defect layer, we simulate the behaviour of transmission modes for a 1D  defective photonic model with the structure $air/(AB)^{N} D (BA)^{N}/air$. Here, $N$ the period of crystal, is chosen to be $N=10$. Furthermore, as recent studies suggest we can ignore the thermal expansion behaviour of these materials in our work [21].}
\par
\emph{As it is illustrated in Fig. 1, our photonic sample is a host of a defect layer which is located at $z=0$ in the middle of two periodically dielectric groups, ($\varepsilon_{1}\:\varepsilon_{2}\:\varepsilon_{1}...$) for $z<0$ and ($\varepsilon_{2}\:\varepsilon_{1}\:\varepsilon_{2}...$) for $z>0$ with an infinite layer of graphene in the $(xy)$ plane placed at the intersection of the defect layer and dielectric groups at both sides. Now, considering the ability of graphene to support both transverse magnetic ($TM$) and transverse electric ($TE$) surface modes and the coupling effect of the magnetic field, the calculation is followed relying on the boundary conditions for normal EM waves penetrating in a graphene layer placed between two dielectric materials. The below equation could show the time-independent electric field ($E_{ix/y}; i=1,2$), as seen in Fig. 1, for  $TM$ polarization with $E_{x}\neq0, E_{z}\neq0, H_{y}\neq0 $ and $TE$ polarization with $E_{y}\neq0, H_{z}\neq0, H_{x}\neq0 $ in both areas based on propagation waves [22-24]:
\begin{equation}
\begin{cases}
E_{1x/y}=a_{1x/y}e^{ik_{1}z}+b_{1x/y}e^{-ik_{1}z}\\
E_{2x/y}=a_{2x/y}e^{ik_{2}z}+b_{2x/y}e^{-ik_{2}z}\\
\end{cases}
\end{equation}
where $i=1$ ($i=2$) is used for $z<0$ ($z>0$). Moreover, $k_{i}=\sqrt{\varepsilon_{i}}\:\frac{\omega}{c}$ represents the wave vector in the two different zones and $\omega$ and $c$ stand for angular frequency  and light speed, respectively. Next, in the presence of magnetic field, the only thing that might attract one's attention is finding out the coupling matrix, $D_{1\rightarrow 2}$, joining the electric fields located at both sides of the defect layer, D which here is $SiC$. Then we write:
\begin{equation}
\begin{bmatrix}
a_{1\pm}\\
b_{1\pm}
\end{bmatrix}
=D_{1\rightarrow 2}
\begin{bmatrix}
a_{2\pm}\\
b_{2\pm}
\end{bmatrix}
\end{equation}
where $a_{i\pm}=a_{ix}\pm ia_{iy}; \quad b_{i\pm}=b_{ix}\pm ib_{iy}$ with the signs $\pm$ related to decoupled right- and left-handed waves, respectively. Now, in order to obtain $D_{1\rightarrow 2}$, it is necessary to write the boundary conditions of the continuity of fields and their derivatives in the connection line at ($z=0$) [25]:
\begin{equation}
\begin{cases}
\textbf{n}\times (\textbf{E}_{2}-\textbf{E}_{1}) \vert_{z=0}=0\\
\frac{\partial\textbf{E}_{1}}{\partial z} \vert_{z=0} -\frac{\partial\textbf{E}_{2}}{\partial z} \vert_{z=0} =i\omega \mu_{0}  \textbf{J} \vert_{z=0}
\end{cases}
\end{equation}
 Here, $\textbf{J}=\sigma \textbf{E}_{2}$ and optical conductivity tensor with matrix elements of $\sigma_{i,j}=$ is the surface current density of the graphene layer. The desired optical conductivity which includes intra- and interband parts is a function of angular frequency ($\omega$), electron scattering rate ($\Gamma$) and of course chemical potential ($\mu_{c}$) written in its tensor form for which the diagonal elements and the off-diagonal illustrate the longitudinal and Hall conductivity are given in [26]. In both above equations, $\pm M_{n}=\sqrt{2nv_{f}^{2}\vert eB\vert\hbar}$ shows different values of discrete LLs. The Fermi distribution function is shown by $f_{d}(\epsilon)=\frac{1}{1+exp(\frac{\epsilon-\mu_{c}}{K_{B}T})}$ in general and $K_{B}$ is the Boltzmann constant. Using the optical conductivity tensor alongside of applying the boundary conditions mentioned earlier, we arrive at four equations which could be written into two groups of expressions for the ease of rapid easy-to-follow illustration as:
\begin{equation}
\begin{cases}
a_{1x/y}+b_{1x/y}=a_{2x/y}+b_{2x/y}\\
k_{1}(a_{1x/y}-b_{1x/y})-k_{2}(a_{2x/y}-b_{2x/y})=\omega\mu_{0}\sigma_{0}(a_{2x/y}+b_{2x/y})\pm\omega\mu_{0}\sigma_{H}(a_{2y/x}+b_{2y/x})
\end{cases}
\end{equation}
The calculations could be followed by multiplying the first expression in (4) by $\pm k_{1}$ and then add it to the second one to write the result correspondingly as:
\begin{equation}
\begin{cases}
2k_{1}a_{1x/y}=k_{1}(a_{2x/y}+b_{2x/y})+k_{2}(a_{2x/y}-b_{2x/y})+\omega\mu_{0}\sigma_{0}(a_{2x/y}+b_{2x/y})\pm\omega\mu_{0}\sigma_{H}(a_{2y/x}+b_{2y/x})\\
-2k_{1}b_{1x/y}=-k_{1}(a_{2x/y}+b_{2x/y})+k_{2}(a_{2x/y}-b_{2x/y})+\omega\mu_{0}\sigma_{0}(a_{2x/y}+b_{2x/y})\pm\omega\mu_{0}\sigma_{H}(a_{2y/x}+b_{2y/x})
\end{cases}
\end{equation}
Coupling now is possible when one multiply $y$-depended case by ($\pm i$) and added to $x$-dependent ones.
\begin{equation}
\begin{cases}
2k_{1}a_{1\pm}=k_{1}(a_{2\pm}+b_{2\pm})+k_{2}(a_{2\pm}-b_{2\pm})+\omega\mu_{0}\sigma_{\mp}(a_{2\pm}+b_{2\pm})\\
-2k_{1}b_{1\pm}=-k_{1}(a_{2\pm}+b_{2\pm})+k_{2}(a_{2\pm}+b_{2\pm})+\omega\mu_{0}\sigma_{\mp}(a_{2\pm}+b_{2\pm})
\end{cases}
\end{equation}
 where $\sigma_{\pm}= \sigma_{x}\pm i \sigma_{y}$. Finally, the transmission matrix ($D_{1\rightarrow 2}$) for two adjacent layers can be read as:
\begin{equation}
\begin{bmatrix}
a_{1\pm}\\
b_{1\pm}
\end{bmatrix}
=\frac{1}{2}
\begin{bmatrix}
1+\frac{k_{2}}{k_{1}}+\frac{\omega\mu_{0}}{k_{1}}\sigma_{\mp}&&
1-\frac{k_{2}}{k_{1}}+\frac{\omega\mu_{0}}{k_{1}}\sigma_{\mp}\\
1-\frac{k_{2}}{k_{1}}-\frac{\omega\mu_{0}}{k_{1}}\sigma_{\mp}&&
1+\frac{k_{2}}{k_{1}}-\frac{\omega\mu_{0}}{k_{1}}\sigma_{\mp}
\end{bmatrix}
\begin{bmatrix}
a_{2\pm}\\
b_{2\pm}
\end{bmatrix}
\end{equation}
It can be expanded to ($m-1\rightarrow m;\;m= 2, 3, 4,... $) in general form where $\sigma_{\mp}$ can be neglected in the absence of graphene layer in order to generalize the above matrix.
}
\par
\emph{Here, a multiplication of transmission matrices across different interfaces is needed to obtain the reflection an transmission of EM waves through 1D PHE system. Defining the propagation matrix ($P$) with $N$ the number of graphene layers, the ($P$) matrix can join the fields at $z+\delta z$ to the fields at $z$ position [27].
\begin{equation}
P(\Delta z)=
\begin{bmatrix}
e^{ik_{z}\Delta z}&&0\\
0&&e^{-ik_{z}\Delta z}
\end{bmatrix}
\end{equation}
\begin{equation}
M=D_{1\rightarrow 2}P(d_{1,2})D_{2\rightarrow 3}P(d_{2,3})....D_{N-1\rightarrow N}P(d_{N,N+1})
\end{equation}
Now, the transmittance ($T$) and reflection ($R$) coefficients can be derived for the ease of numerical evaluations as follows.
\begin{equation}
R= \vert\frac{M_{21}}{M_{11}}\vert^{2}\; ;\qquad T=\vert\frac{1}{M_{11}}\vert^{2}
\end{equation}
}
\emph{Numerical simulations, then, based on the above relations can be carried out to reveal the defective transport spectra of the 1D dedective structure. In Fig. 2, the transmission spectra of the EM modes for right-handed polarized propagation are shown. It is seen that defective modes with larger transmission at stronger fields are appeared in the forbidden frequency range of the photonic structure corresponding to the different values of the applied magnetic field, $B=1$, $5$, $10$, $20$ and $40$ Tesla. It is also clear that the position of the emerging modes are also affected such that the flow of the modes within the considered frequency range for our simulation is toward lower region of the far-infrared spectrum. The similar situation seems to be true for the left-handed polarized defect states. Fig. 3 shows the result for defective transmission of left-handed modes within the same frequency zone. In this case we also see that the transmission shows an increase at stronger magnetic fields. However, unlike circularly right-handed defect modes, correspondingly, some left-handed states seem to be diminished. To be more specific, for magnetic fields $5$ and $10$ Tesla, almost, no defective modes emerge and the appearing modes relative to the right handed case give lower transmission. Another interesting feature of these left-handed states is that for magnetic fields changing from $1$ to $20$ Tesla, the flow in position toward lower parts of the photon spectrum exhibits a rather far modulation comparing to the results obtained in Fig. 2. Remarkably, however, in this case, the transmission spectra are more affected outside the forbidden frequency range. We also note that the high transmitted mode emerging at relatively stronger magnetic field, $B=40$ Tesla, is blue-shifted. This transitional shifting of defect modes in position with increasing the magnetic field which also could root to LLs need to be more analyzed since we know that the optical conductivity which itself is reflected by transmission of the electromagnetic defective waves, reflects the LLs pattern in graphene. Therefore, we can seek the optical QHE system by simulating transmission as a function of varying the magnetic field at low temperatures and stronger fields.}
 \par
 \emph{Fig. 4 shows the simulation for defective spectra of right-handed states at very low temperature, $T=1\ K$ as a function of the magnetic field for a frequency interval $22.25$ to $22.65$ $meV$. Surprisingly, we see that emerging modes appear in a plateau-like structure resembling the well known feature seen for QHE in 2D electronic systems which we refer them to as quantum Hall defect modes (QHD modes). The transmission plateaus for circularly polarized electromagnetic defect modes show that increasing the magnetic field leads to higher transmission for QHDs. Interestingly, however, at stronger magnetic fields ( larger than $30\ Tesla$ ) an oscillatory behaviour for QHDs is observed (see Fig. 4). This situation would be better illustrated if we go to frequency space and show the transmission spectra as a function of the photon frequency as shown in Fig. 5. Here, numerical simulations show that appearing QHDs are more localized at stronger magnetic field regime. In this relatively high field domain, the lowest density of defect states corresponds to oscillatory transmission of QHDs. In fact, one can say that these high transmitted modes strongly remain localized and oscillate in their position. In contrast, QHDs emerging at very low field regime are strongly delocalized, creeping from higher frequencies to lower parts where they begin to be more localized which shows why the width of plateaus increases with the magnetic field. As it clear in Fig. 5 the flow of right-handed QHDs is toward more localization in lower frequencies at stronger fields. }
 \par
 \emph{Let us now study this transmission feature for left-handed QHDs by simulating the effect of magnetic fields from low regime to stronger domains for frequency interval as shown in Fig. 6. As it observed the dominant step-like structure is emerged for magnetic fields ranging from 15 to 45 Tesla which correspond to the higher transmitted modes observed in Fig. 3. Moreover, one can see that the largest transmission leap occurs around $B=30$ Tesla for which the oscillatory plateau is observed. Here, we also examined the effect of increasing the temperature on the step-like spectra in the same frequency range and the magnetic field domain. Obviously, in the case of increasing temperature to $T=100\ K$, it is observed that quantum Hall defective transitions tend to vanish. Note that in Fig. 6 for $T=100\ K$ (blue) and $T=10\ K$ (red) the transmission around $B=20$ Tesla for which the two plots almost intersect at the middle of the corresponding plateau is the same which means that upon diminishing of transmission plateaus the increasing temperature provides two region before te intersecting point; one with lower and another with higher transmission relative to the plateau. We see that this is also the case for right-handed QHDs if we turn our attention back to these modes and examine the effect of temperature on the corresponding behaviour.  This effect is illustrated in more details in Fig. 7 where the defective transmission as a function of the increasing magnetic fields ranging from $1$ to $45$ Tesla for three different temperatures $T=10\ K$, $T=100\ K$ and $T=300\ K$ are shown. We see more significant transitional behaviour with decreasing in width at low fields limit. From this figure it is also obvious that the intersecting point shifts to higher fields as the temperature increases. To see the oscillatory transmission step by increasing the temperature we focus on QHDs behaviour in the frequency domain under increasing the temperature. Therefore, we plot the situation, as shown in Fig. 8, for frequency ranging from $22.3$ to $22.65\ meV$ where the right-handed QHDs emerge for $T=10\ K$ (red) $T=300\ K$ (blue). It shows that how increasing temperature spread the localized modes along the whole frequency region. It might now then be qualitatively understood why the oscillatory step tend to dimmish at high temperatures. Comparing with Fig. 5 shows that for $T=10\ K$ more excitations is observed and, therefore, one can expect at zero temperature completely localized states. Now, As it is seen in Fig. 9 for left-handed modes, the flow of QHDs in position is toward regions of higher frequencies upon increasing the external field for frequency ranging from $22$ to $22.25\ meV$ where the dominant left-handed QHDs emerge (see also Fig. 3).
 }
 \par
 \emph{Finally, for further study of the properties of step-like pattern of emerging QHDs, we choose to examine the influence of the chemical potential on the spectra for right-handed states which up to this stage was considered to be $\mu=0.2\ eV$. At $T=10\ K$, as it is clear from Fig. 10, increasing chemical potential from $0.2\ eV$ to $0.6\ eV$ in the same field domain changing from $15$ to $45$ Tesla reduces the transmission and split the plateaus into more steps.
 }
\section{Conclusion remarks}\label{section.five}
 \emph{Based on the transmittance properties of the electromagnetic waves in photonic structures and the peculiar optical properties of graphene, a photonic quantum Hall system was introduced [30-32]. In this work, just focusing on the main ideas, a 1D PHE system, by emerging quantum Hall defective EM modes for a graphene-covered 1D defective optical structure was marked. It was theoretically by simulating the behaviour of defective spectrum, demonstrated that the transmission of electromagnetic modes shows transmission plateaus which is an important well known feature observed earlier for conductance of 2D electronic systems under the influence of normal external magnetic field. The main framework of our work was based on IQHE in gapless graphene with low loss propagation of EM waves. On the analytical side, we used the transfer matrix method to derive the transmission coefficients for two decoupled EM modes under the effect of external magnetic field. The behavior of EM waves in 1D structure was considered by commercial simulations which smoothen our way toward detecting Hall defective structure of the system. Significantly, the transmission steps were emerged within a relatively narrow frequency interval which resemble the static electronic quantum Hall effect which occur for zero frequency. }
\par
  \emph{In our analysis, we first discussed the transmission steps as a function of the applied magnetic field for two circularly polarized transport. The most significant feature to be noted in this case for QHDs was growing the transmission with growing both the height and width of the steps simultaneously with the localization in the position for increasing the intensity of the magnetic field. Such behaviour also represented a magneto-transmission oscillation for QHDs at the largest transmission leap which has no electronic counterpart in 2D electronic gas systems. Analyzing the situation in frequency space revealed that these oscillatory states which provide also an interesting background for further invitations were strongly localized in the position. Apart from similarities that right and left-handed QHDs exhibited, we observed that a notable difference was that higher transmission modes for higher magnetic fields would occur for the flow of right-handed QHDs toward lower frequencies while gaining the significant growing transmission with increasing the magnitic bias for left-handed modes would be the case toward upper parts of the forbidden frequency region.  }
   \par
 \emph{ Inspired by the role of the temperature in electronic QHE systems, we also examined the effect of increasing the temperature on the QHDs. Since we were interested in the overall effect we merely considered some special situations which revealed that the leaps in transmission for both right and left-handed QHDs would be replaced with a transmission curve. The delocalization and, therefore, reducing the density of the modes in the frequency line without very significant change in the transmission was another significant result to be noted in this case. Further, oscillations of the Hall defect states appearing in the last step diminished with increasing the temperature. This domain of sensitivity would open a subject of interest for optical thermal detectors in real world applications.  }
 \par
   \emph{Apart from magneto-tunable situations for QHDs, the tunability properties of the quantum Hall polarized states for some interested situations in THz region for a fixed domain of the magnetic field and the frequency was also discussed. Tuning the chemical potential in a constant magnetic background for right-handed modes as our choice for our investigation in this case revealed that effect of doping create splitting in the transmission steps while decreasing it. Accepting the role of the localization in position in broadening the transmission steps in the magnetic line, the delocalization would therefore spread the modes in the frequency line upon increasing the chemical potential. Finally it should be emphasized that, our work was intended to be a rapid communication and therefore the main general results was just tried to be covered by commercial simulations. Other details regarding the dielectrics, the change in structure of the sample as so on which is beyond the aim of this work can be carried out both theoretically and experimentally. }
   \par
   \emph{To summarize, using a 1D defective photonic structure, the behaviour of optical transmission spectrum of massless Dirac fermions of graphene under a constant magnetic field was investigated in THz region. The existence of a photonic Hall effect by emergence of optical defective modes with rich optical properties was demonstrated. It was found that increasing the magnetic field suggests highly tunable modes with growing in the transmission within a relatively narrow frequency domain which could be simply tested by experiments.}
}
\section{Acknowledgement}{\emph{D. Jahani initiated the study and contributed to the all theoretical calculations, analysis and writing the paper and as an independent researcher invited A. Alidoust, L. Abaspour and T. Jahani to contribute in checking the simulations, the text and the final overall calculations check of the manuscript, respectively. }}

\begin{figure}
 \begin{center}
\includegraphics[width=10cm]{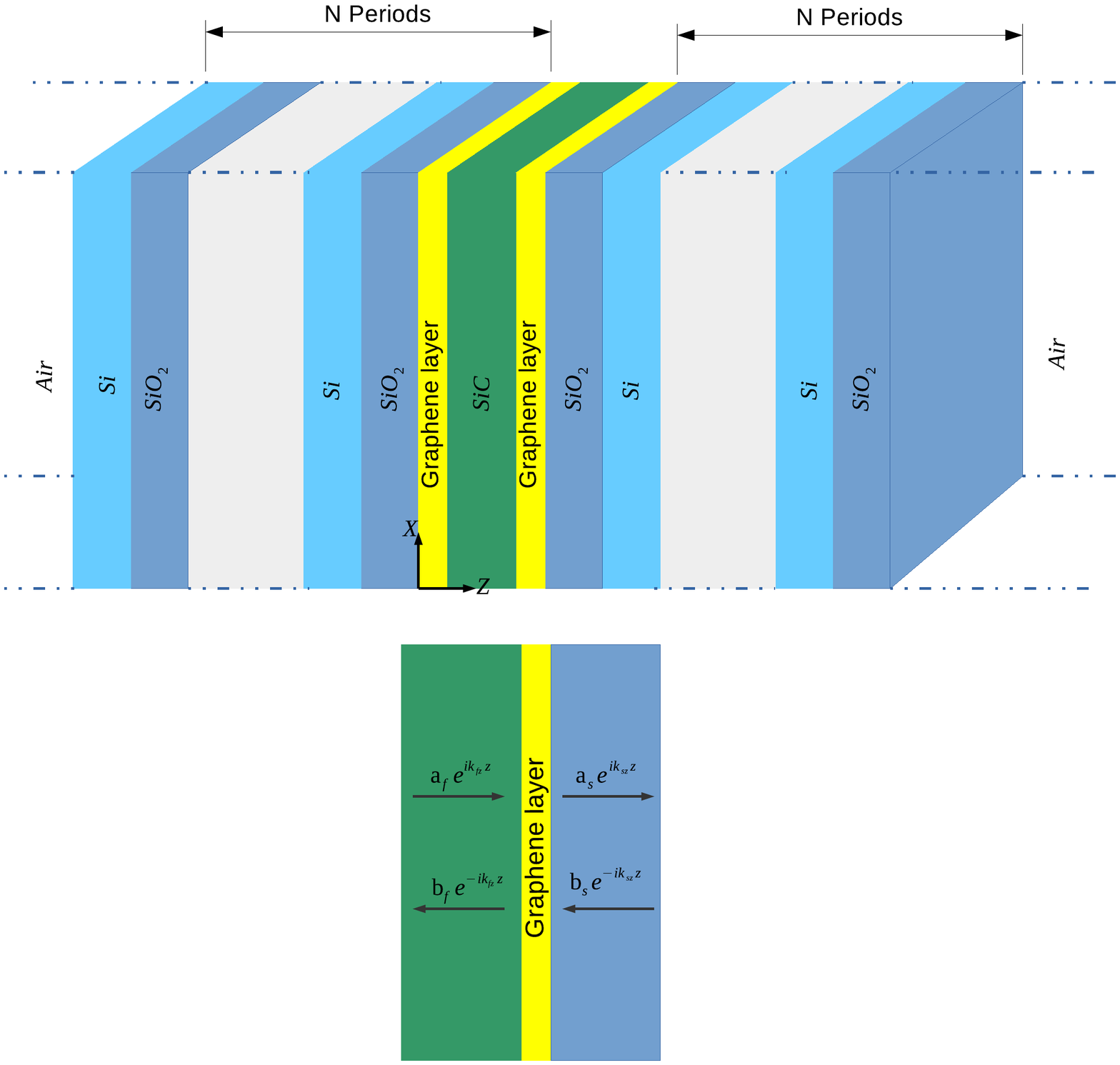}
\caption{Left: A stack of two N- periods of different dielectrics placed at both sides of the graphene-covered
defect layer drawn alongside of z axis is shown. Right: a zoomed-in graph of two adjacent layers where
graphene is placed between them is enlarged for which the time-independent electric field components are
shown.}
\end{center}
\end{figure}
 \begin{figure}
 \begin{center}
\includegraphics[width=18cm]{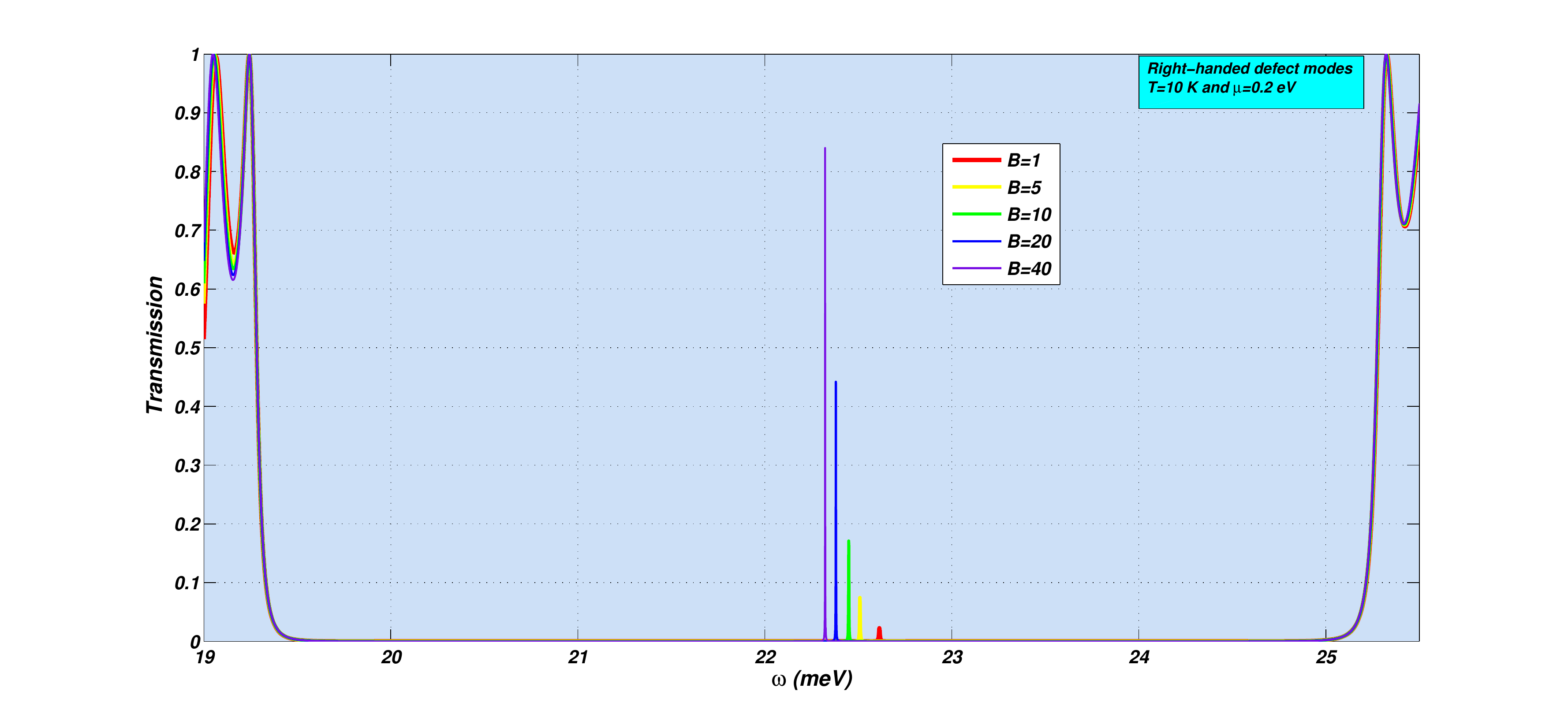}
\caption{Schematic representation of the transmission of QHDs in the PBG of
the proposed optical system for right-handed polarized states at $T = 10\ K$ and $\mu = 0.2\ eV$. The increasing the magnetic field shows overall red-shifting of QHDs in a narrow range of the frequency space within the PBG zone of around $19$ to $25.5\ meV$. Out of PBG zone it seems that the transmission is more affect relative to the effect observed for right-handed QHDs. }
\end{center}
\end{figure}
\begin{figure}
 \begin{center}
\includegraphics[width=18cm]{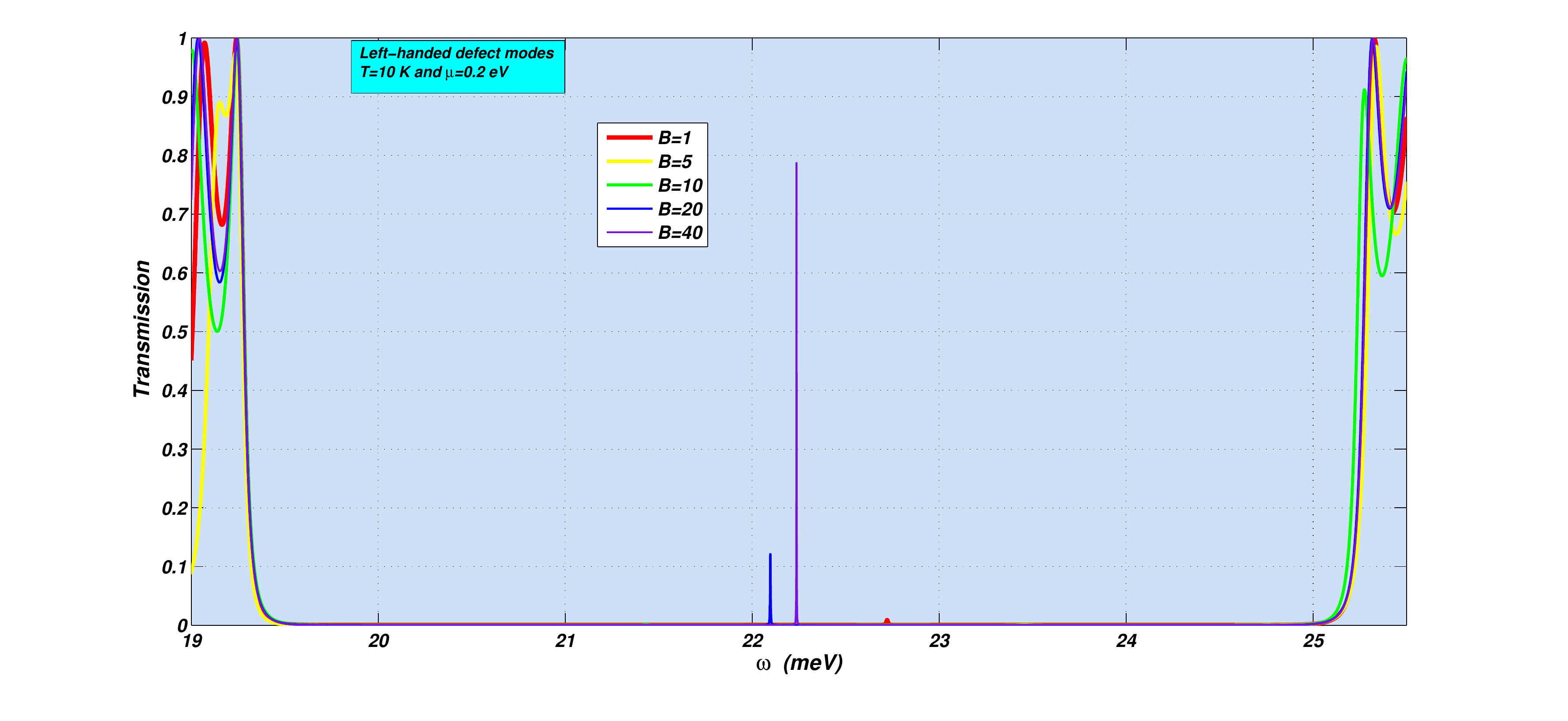}
\caption{The illustration of effect of increasing the magnetic field on the transmission in the PBG of our
sample for left-handed polarized states at fixed values for the temperature and the chemical potential which results in filtering some
modes observed in right-hand case. We see changing from red-shifting to the blue-shifting upon applying a magnetic field $40\ $ Tesla in this case.   }
\end{center}
\end{figure}
\begin{figure}
 \begin{center}
\includegraphics[width=18cm]{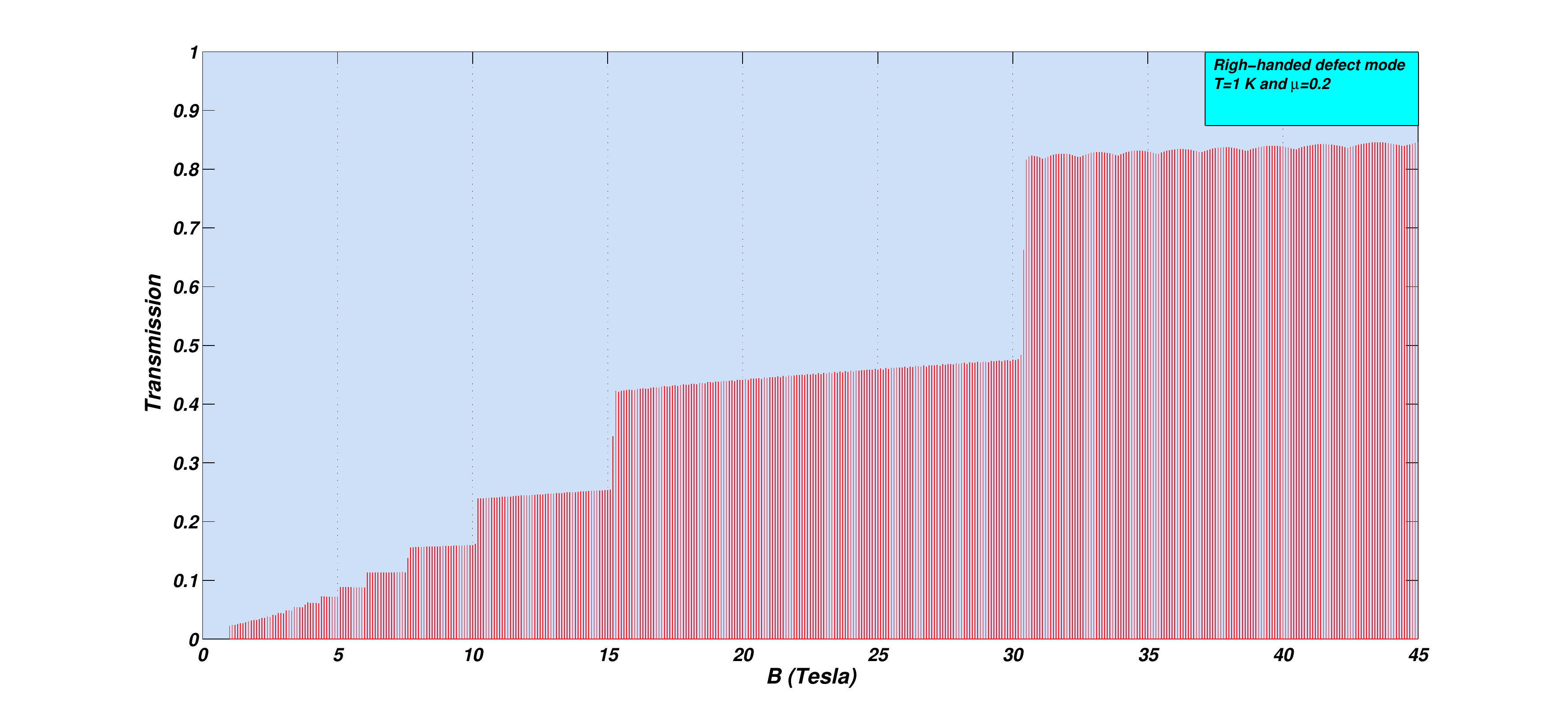}
\caption{PHE transmission spectra of right-handed QHDs as a function of the magnetic field. It is observed that a cost to be
paid for higher transmission, is broadening the range of magnetic field interval. This situation is also seen
for conductance in QHE regime related to 2D electron systems. Special attention should be paid to the oscillatory
behavior for QHDs at stronger magnetic fields in the last step which has not been observed for the electronic Hall effect situation.  }
\end{center}
\end{figure}
\begin{figure}
 \begin{center}
\includegraphics[width=18cm]{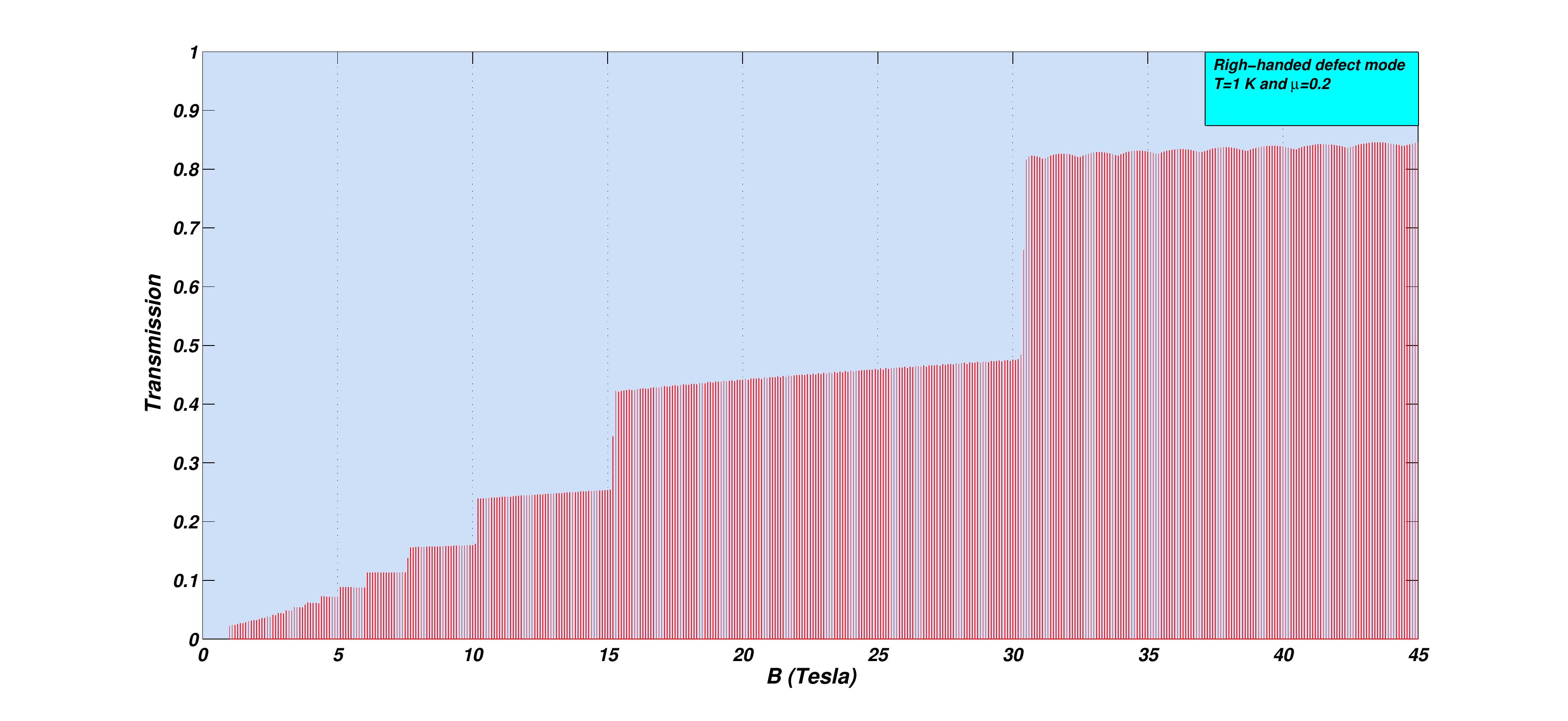}
\caption{PHE transmission spectra for right-handed QHDs for applied magnetic fields
ranging from $1$ to $ 45$ Tesla in the frequency space. It seems lower frequencies shows higher transmission modes with more
localizing effect. As it is clear, the effect of stronger magnetic fields is, therefore, red-shifting the photonic plateaus in Fig. 4.  }
\end{center}
\end{figure}
\begin{figure}
 \begin{center}
\includegraphics[width=18cm]{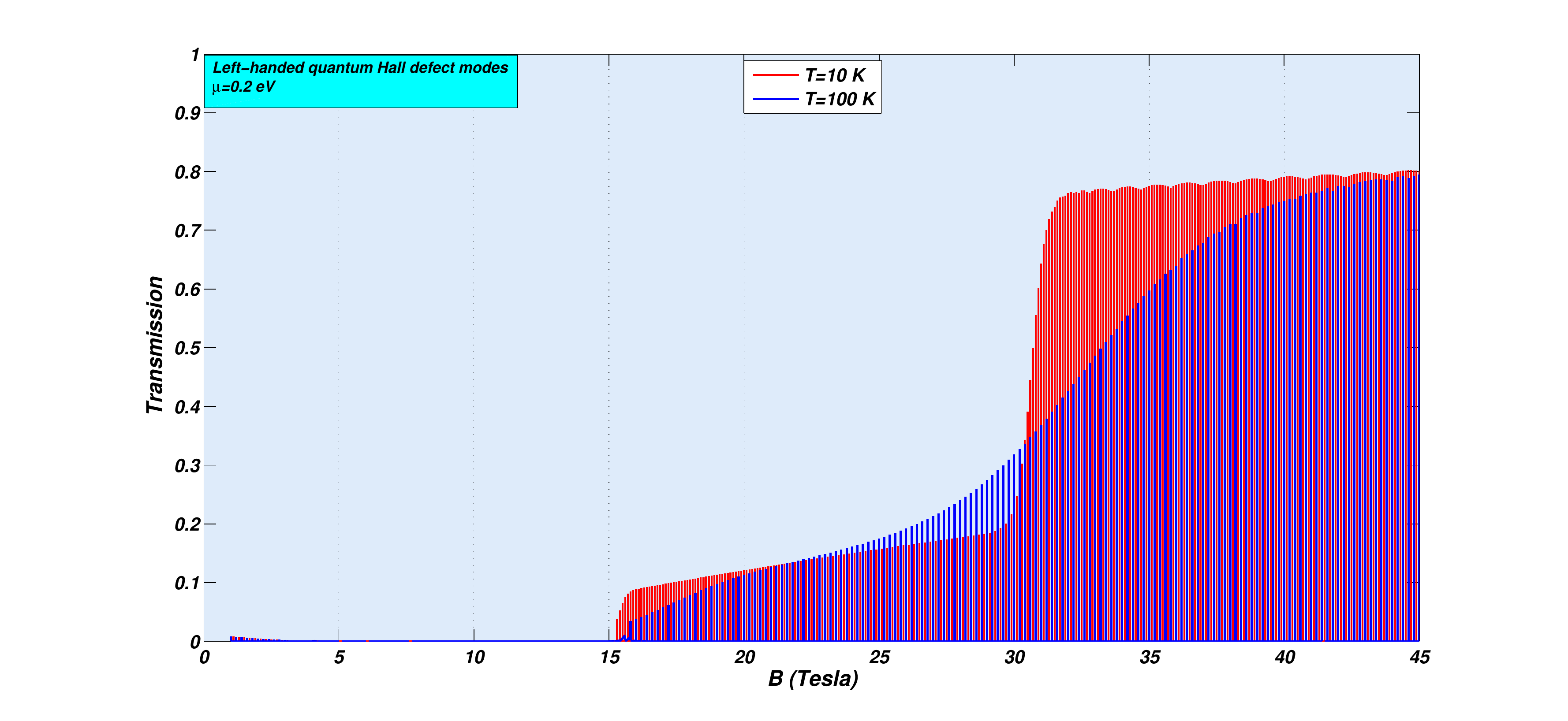}
\caption{ The effect of increasing the temperature on PHE for left-handed
QHDs as a function of the magnetic field ranging from $1$ to $ 45$ Tesla. PHE for left-handed QHDs is almost disregarded upon higher temperatures. }
\end{center}
\end{figure}
\begin{figure}
 \begin{center}
\includegraphics[width=18cm]{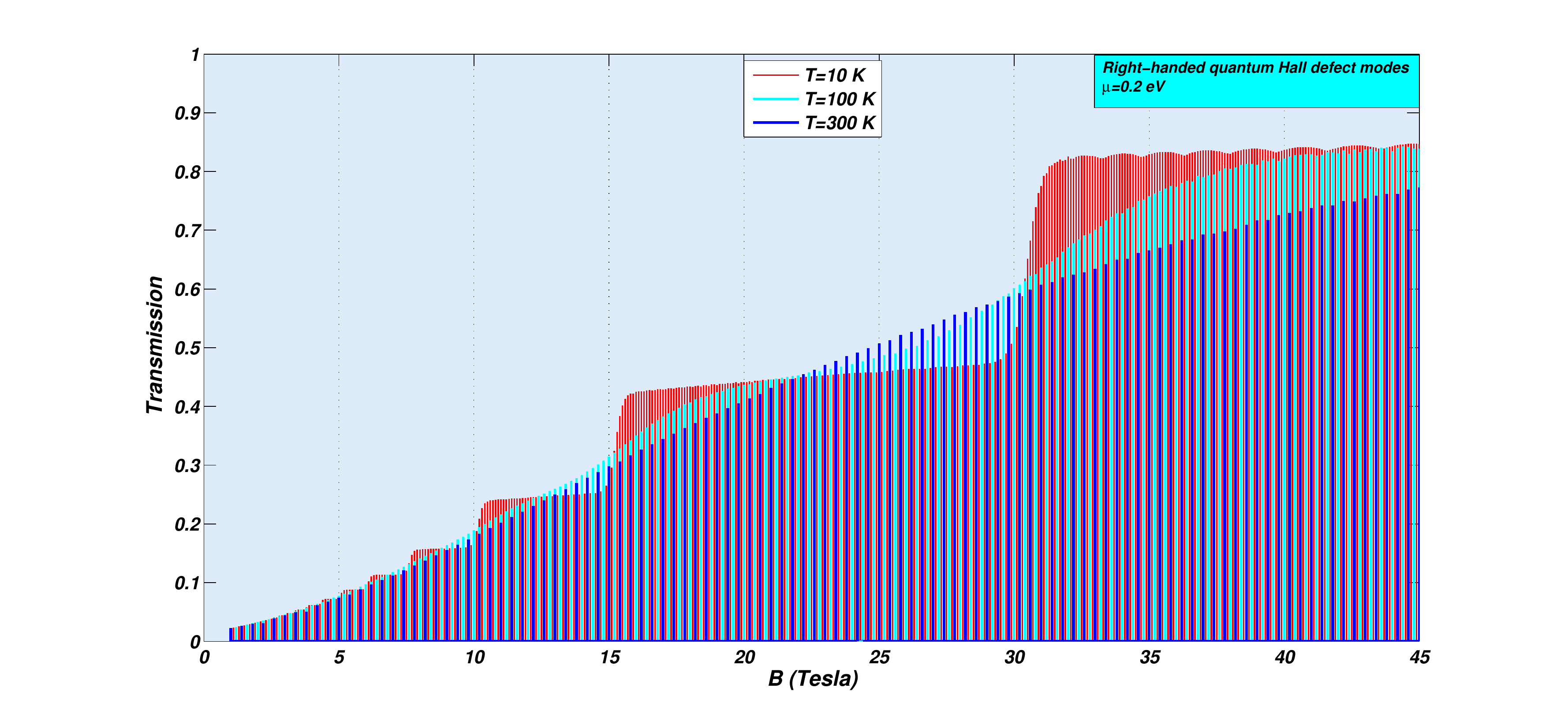}
\caption{The effect of increasing the temperature on PHE for right-handed
QHDs as a function of the magnetic field ranging from $1$ to $ 45$ Tesla. As it is also expected the step-like scheme of the transmission is vanished at higher temperatures. }
\end{center}
\end{figure}
\begin{figure}
 \begin{center}
\includegraphics[width=18cm]{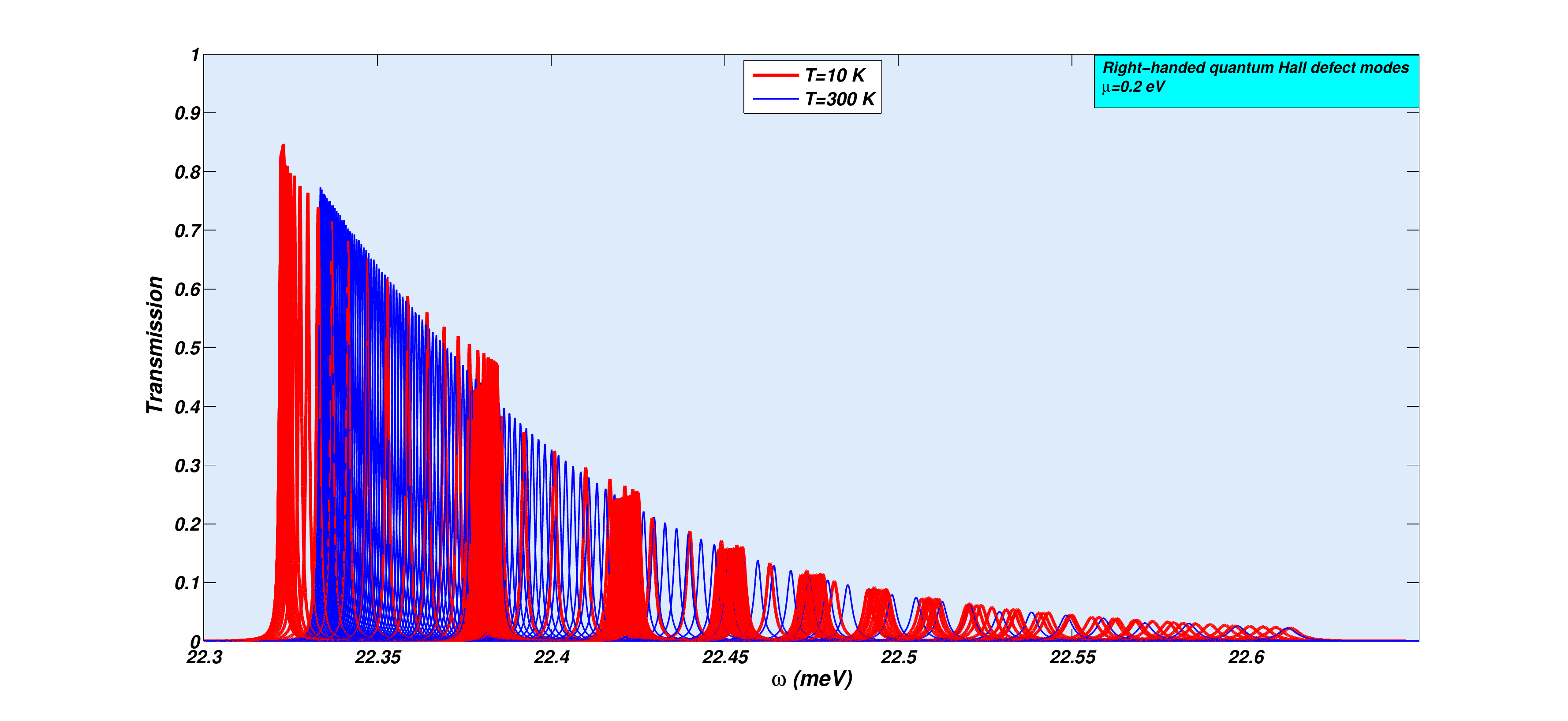}
\caption{ The clear effect of higher temperatures on delocalizing the right handed modes in the frequency space for which, by comparison with figure 4, at $T = 10\ K$ we see that broadening of the QHDs are observed in the frequency domain.  }
\end{center}
\end{figure}
\begin{figure}
 \begin{center}
\includegraphics[width=18cm]{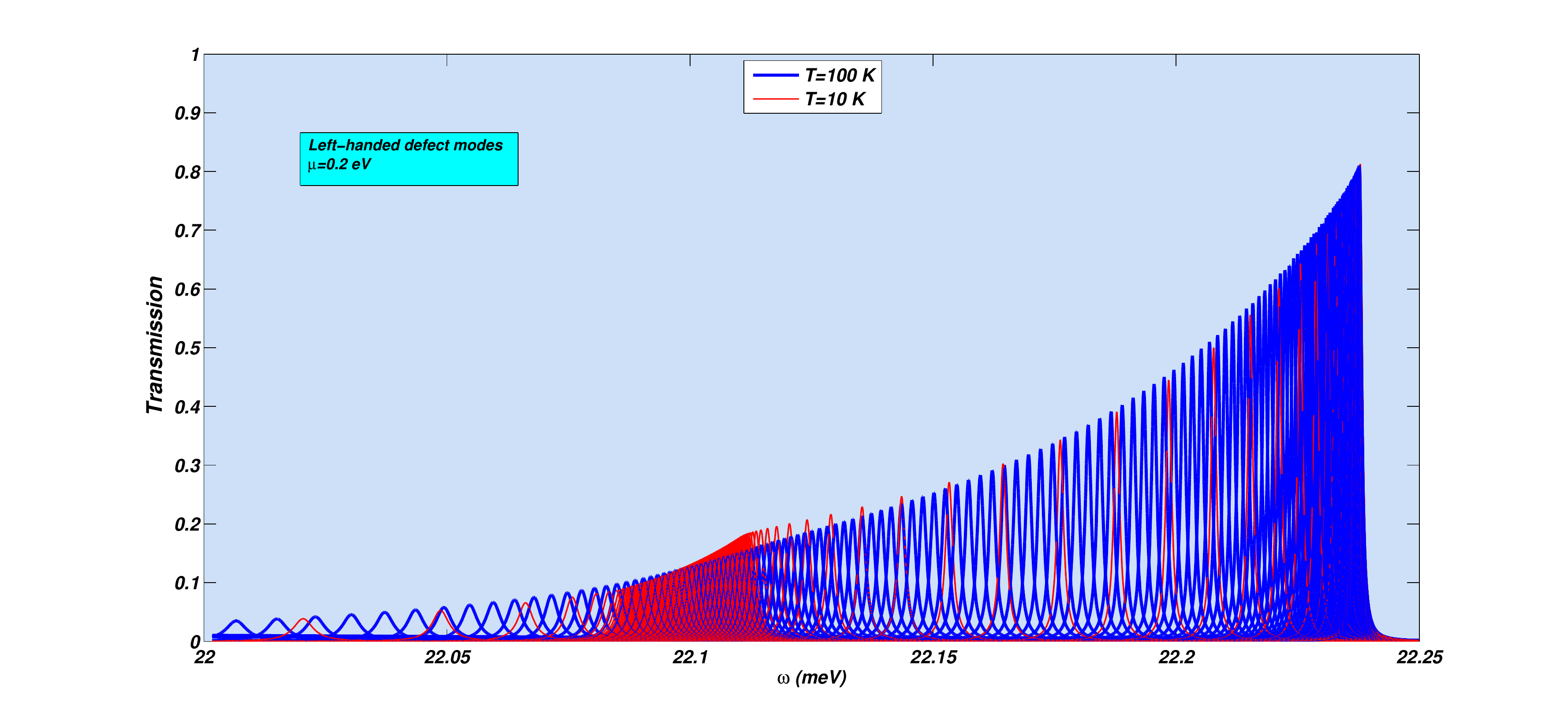}
\caption{ The dependence of transmission of left-handed modes on the magnetic field changing from $1$ to $20$ Tesla for two different temperatures is shown in the frequency domain.  As we see the transmission modes upon both red-and blue-shifting get more localized by decreasing the temperature.    }
\end{center}
\end{figure}
\begin{figure}
 \begin{center}
\includegraphics[width=18cm]{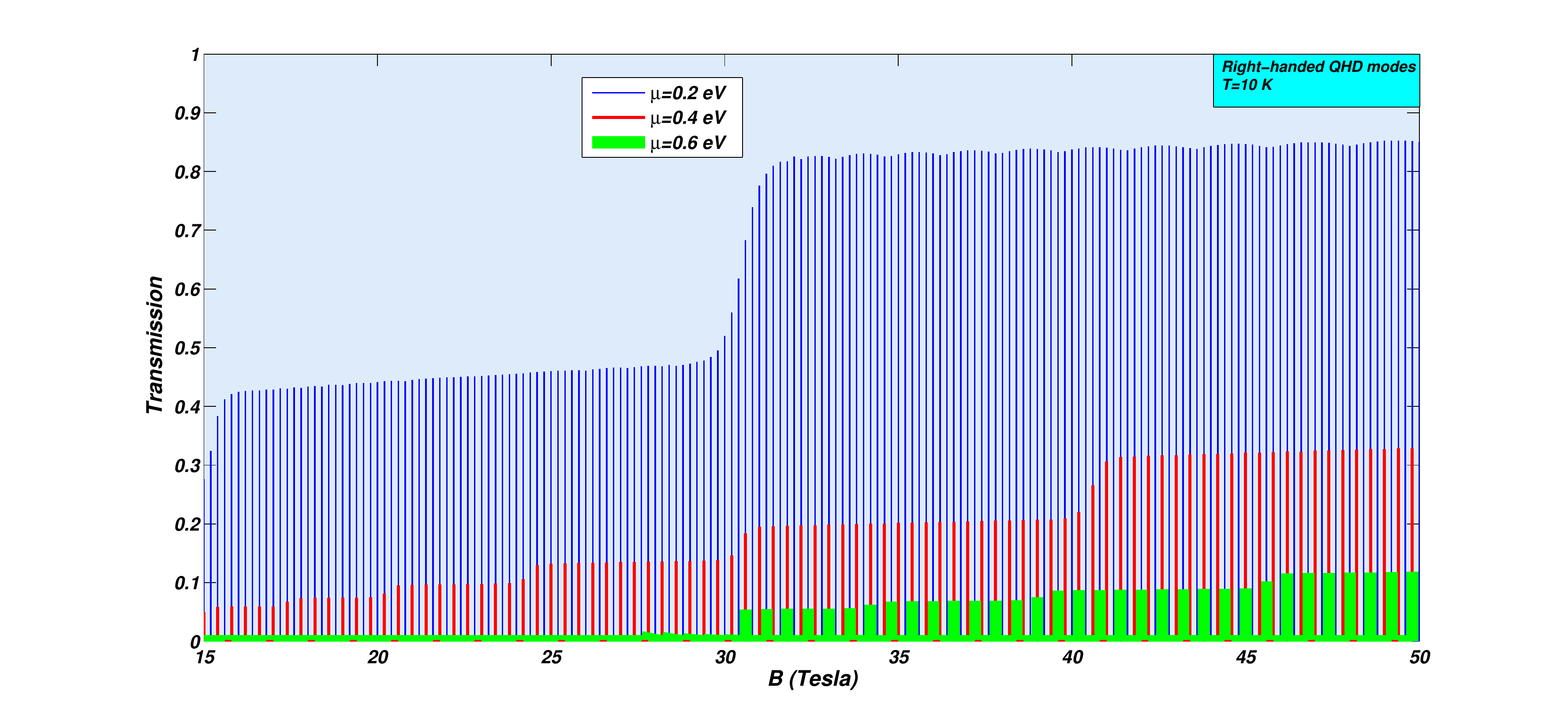}
\caption{The effect of chemical potential on the properties of the QHDs pattern for the
magnetic field ranging from $15$ to $50$ Tesla. More step-like spectrum occurs as the chemical potential
increases. The reduction of the transmission map is also another effect of increasing of the chemical potentia.}
\end{center}
\end{figure}
\end{document}